\documentclass[twocolumn,showpacs,prb,amsmath]{revtex4}
\usepackage{bm}
\usepackage{epsfig}
\usepackage{ulem}
\newcommand{\nix}[1]{}
\usepackage{color}
\usepackage{upgreek}

\begin{document}

\title{Spin and orbital mechanisms of the magneto-gyrotropic photogalvanic effects 
in GaAs/AlGaAs quantum well structures\\
}
\author{ V.~Lechner,$^1$ L.E.~Golub,$^2$  F.~Lomakina,$^{1}$ V.V.~Bel'kov,$^{2}$ P.~Olbrich,$^1$ S.~Stachel,$^1$ I.~Caspers,$^1$
M.~Griesbeck,$^1$ M.~Kugler,$^1$ M.J.~Hirmer,$^1$ T.~Korn,$^{1}$ C.~Sch{\"u}ller,$^{1}$ D.~Schuh,$^1$ W.~Wegscheider,$^{3}$
and S.D.~Ganichev$^{1}$
}
\affiliation{$^1$Terahertz Center, University of Regensburg,
93040 Regensburg, Germany}
\affiliation{$^2$Ioffe Physical-Technical Institute, Russian
Academy of Sciences, 194021 St.\,Petersburg, Russia}
\affiliation{$^3$ETH Zurich,
8093 Zurich, Switzerland}

\begin{abstract}

We report on the study of the linear and circular magneto-gyrotropic 
photogalvanic effect (MPGE) in GaAs/AlGaAs quantum well structures. 
Using the fact that 
in such structures
the Land\'{e}-factor $g^*$ depends on the 
quantum well (QW) width and has different signs for narrow and wide QWs, 
we succeeded to separate spin and orbital contributions to both MPGEs. 
Our experiments show that, for most quantum well widths, the MPGEs are  mainly 
driven by spin-related mechanisms, which results in a photocurrent 
proportional to the $g^*$ factor. In structures with a vanishingly 
small $g^*$ factor, however, linear and circular MPGE are also detected, proving the 
existence of orbital mechanisms.

\end{abstract}

\maketitle

\section{Introduction}

Spin and orbital mechanisms, which are present simultaneously in various physical phenomena, result 
in two competitive contributions in the observable effects. Textbook examples therefore are 
Pauli paramagnetism and Landau diamagnetism that yield two comparable contributions to the 
magnetic susceptibility of an electron gas.~\cite{LL5} Another bright manifestation is the fine 
structure of  exciton lines in a magnetic field.~\cite{Exc} In all these cases, the electron systems are affected by a magnetic field 
in two different ways: Via Zeeman splitting of spin sublevels and due to 
cyclotron twisting of electron trajectories.
An interplay of spin and orbital mechanisms is also expected for
linear and circular magneto-gyrotropic photogalvanic effects (MPGE). 
The spin-related mechanisms of both MPGEs were widely discussed in 
the past and are driven by the spin-dependent relaxation of a non-equilibrium electron gas 
in gyrotropic two-dimensional electron systems (2DES)
(for reviews see Refs.~\onlinecite{Fabian08,Ivchenkobook2,GanichevPrettl,Winkler06,Ivchenko08,SSTreview}).
The microscopic mechanisms of the spin-driven MPGEs are based on spin-orbit coupling 
in 2DES with structure  and bulk inversion asymmetry (SIA and BIA). 
In  the case of the  linear MPGE the absorption of linearly polarized or unpolarized 
radiation leads to electron gas heating. Due to 
a spin-dependent energy relaxation of these heated electrons, two equal and 
oppositely directed electron flows in the
spin-up and spin-down subbands result, what represents  a pure spin current.~\cite{Ivchenko08,SSTreview,Handbook,naturephysics06,condmat110} 
The Zeeman splitting of the subbands induced by the in-plane magnetic field leads to
the conversion of a spin flow into a measurable spin-polarized electric current.
The circular MPGE yields a photocurrent, proportional to the
radiation helicity. It is caused by the spin-galvanic effect (SGE),~\cite{Nature02,SGEopt} 
in which the spin-flip relaxation of a
spin-polarized non-equilibrium electron gas results in an electric current.~\cite{Ivchenko08,SSTreview,review2003spin} 
Here, an in-plane magnetic field rotates the optically induced spin polarization into the plane of the 2DES due to the Larmor
precession. This provides an in-plane spin component, which is necessary for the SGE in (001)-grown QWs.
While the spin-based origin of the MPGE was intensively  discussed,
most recently it was pointed out
that orbital effects, caused by a magnetic-field-induced scattering asymmetry,
may substantially contribute to the MPGE~\cite{Tarasenko_orbital,Tarasenko_orbitalMPGE2} and thus complicate the analysis 
of the spin currents. The latter is of importance for the exploration of 
spin generation and  spin-related transport in 2DES, which 
are  major and still growing fields in solid-state 
research.~\cite{Fabian08,Dyakonovbook,Awschalombook2010,Wu2010}

Here we report on experiments, which allow us to distinguish unambiguously 
between the spin-dependent and orbital origin of the MPGE by
investigating the qualitative difference in their behavior upon a
variation of the $g^*$ factor. We use the fact that the
electric current resulting from the spin-related roots is proportional
to the Zeeman band spin splitting in contrast to the one due to 
orbital mechanisms. To explore this difference, we utilize the well
known fact that in GaAs based quantum wells the Zeeman splitting
changes its sign  at a certain QW width.~\cite{Snelling91,Ivchenko_g_factor,yugova} 
This  inversion is mostly caused by 
the opposite signs of the $g^*$ factor in the 
GaAs  with respect to the
AlGaAs-barrier and the fact that for narrow QWs the electron wave
function deeply penetrates into the barrier. Therefore, we studied
the MPGE in a set of 
structures with QW widths varying from 4 to 30~nm.

The paper is organized as follows. In Sec.~\ref{technique} we describe details 
of the sample preparation and present an overview of the experimental 
techniques used to study the MPGE. In Sec.~\ref{kerr} we give 
an account on combined photoluminescence (PL) and PL 
excitation (PLE) measurements as well as time-resolved Kerr 
rotation (TRKR) experiments performed on the sample series in order 
to determine the electron confinement energies and the electron 
$g^*$ factor, respectively. Sections~\ref{LMPGE} and~\ref{CMPGE} deal with the experimental study 
and  analysis of the
linear and circular MPGE, respectively. Finally, we summarize the study.

\section{Samples and experimental techniques}
\label{technique}

\begin{figure}[ht]
\includegraphics[width=0.45\textwidth]{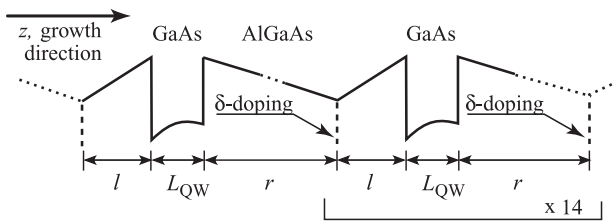}
\caption{Conduction band structure of the samples.}\label{fig1}
\end{figure}

The experiments were carried out on (001)-oriented
Si-$\delta$-do\-ped $n$-type  GaAs$/$Al$_{0.3}$Ga$_{0.7}$As  structures grown by
mo\-le\-cu\-lar-beam epitaxy at typical temperatures in excess of 600$^\circ$C.
A set of structures  with the same doping profile but
different QW widths $L_{\rm QW}$,  between  4~nm and 30~nm,
was grown. Figure~\ref{fig1} sketches the conduction band  edge 
of the multiple QW structures together with the corresponding 
$\delta$-doping positions, and Table~\ref{samples} gives the sample parameters. 
The doping layers are asymmetrically shifted
by two different spacer layer thicknesses,
$l$ and $r$, with respect to the QW.
In addition, in all samples, except A,
the sum $(l + r + L_{\rm QW})$ was kept constant.
The impurities' Coulomb field yields an asymmetric
potential profile inside the QWs causing SIA.
The investigated samples were square shaped ($5 \! \times \! 5 \:
{\rm mm}^2$) and its edges oriented along the $ [1{\bar 1}0]$ and
$[110]$ crystallographic directions. For photocurrent measurements,
ohmic contacts were alloyed on the sample corners allowing to
probe the photocurrent along the $x \parallel 
[100]$ and $y \parallel [010]$ crystal axes.

The MPGE is measured at room temperature by exciting the samples with linearly 
or circularly polarized te\-ra\-hertz (THz) laser radiation under normal incidence.
As radiation source we used an optically pumped NH$_3$ molecular laser~\cite{GanichevPrettl,JETP1982}
operating at a wavelength $\lambda$ of $280\,\upmu$m. 
The wavelength of $280\,\upmu$m (photon energy of 4.2~meV) was chosen to cause 
free carrier absorption only. 
The laser provided single pulses with a pulse duration of about 100~ns and
a peak power $P$ of about $6$~kW.
The photocurrent $\bm{J}$ is measured via the voltage drop across 
a 50~$\Omega$ load resistor with a digital oscilloscope.
To study the linear MPGE we applied linearly polarized radiation with the radiation 
field $\bm{E}~\parallel~x$, and an external in-plane magnetic field of $B_y$ of $\pm~1$~T.
The resulting current
is measured along the $x$-direction, normally to the magnetic field, $B_y$. 
To excite the circular MPGE we used elliptically  polarized radiation obtained by
a crystal quartz quarter-wave plate. The helicity of the radiation 
was varied by the rotation of the plate according to $P_{\rm circ} = \sin 2\varphi$, where $\varphi$ is the angle between the initial plane of
polarization and the optical axis of the $\lambda/4$-plate. 
In this case the photocurrent $J_x$ is measured in the direction parallel to $B_x$. 
The experimental geometries with  $J_x \perp B_y$ for linear and  $J_x \parallel B_x$ for the
circular MPGE are chosen to probe the currents 
solely caused  by the structure inversion asymmetry.~\cite{BIASIAPRB,LechV12}
This allows us to reduce the influence of the QW width on the degree of
asymmetry. For any other configuration we would also obtain a BIA-induced MPGE,
which complicates the data analysis, because BIA itself strongly depends on $L_{\rm QW}$.~\cite{Harley07}

\section{Sample characterization by PL and TRKR techniques}
\label{kerr}

\subsection{Experimental technique}

To study Land\'{e} factors and spin dynamics we used the time-resolved Kerr rotation  and photoluminescence
techniques, applying a pulsed Ti-Sapphire laser system. 

\begin{widetext}
\begin{center}
\begin{table}[hbt]
\caption{Sample Parameters.
Carrier density $n_s$ (per QW-layer), mobility $\mu$,
momentum scattering time $\tau_p$ calculated from the mobility,
measured  $g^*$ factor and energy difference $\Delta E$ between the
transitions (e2hh2)-(e1hh1) are given for $T=4.2$~K. Also listed: Carrier density $n_s$(RT) (per QW-layer)
and mobility $\mu$(RT) measured at room temperature.
}
\begin{tabular}{ccccccccccccc}
\hline %
\hline
Sample & \:$L_{\rm QW}$ & \: Spacer $l$  & \: Spacer $r$ \: &   \:  $\chi=$   \:     & \: $n_s$  \:      & \: $\mu$  \: 
                        & \: $\tau_p$ \: & \: \: $g^*$ \: & \: $\Delta E$ \:  & \: $n_s$ (RT) \:     & \: $\mu$ (RT) \:  \\
       &   [nm]         &  \: [nm]       &      [nm]        & \:$\frac{l-r}{l+r}$\:  &   \:$10^{11}$[$\frac{\rm{1}}{\rm{cm}^2}$]\:  &   \:$10^5$   
 [$\frac{\rm{cm}^2}{\rm{Vs}}$]\:  & \: [ps] \:     & \:    & \: [meV] \:       & \:$10^{11}$[$\frac{\rm{1}}{\rm{cm}^2}$]\:  
     &  \:$10^3$ [$\frac{\rm{cm}^2}{\rm{Vs}}$]\:  \\[0,07cm]
\hline
  A \:  & 4   & \:  45   &  140  & -0.51   & 1.17  & 0.509 & 1.9  & \: 0.13   & -     & 1.4   & 4.8     \\ \hline
  %
  B  \: & 6   & \: 74.5  & 119.5 & -0.23   & 0.874 & 0.925 & 3.5  & \: 0.106  & -     & 1.3   & 5.5     \\ \hline
  %
  C \:  & 8.2 & \: 73.4  & 118.4 & -0.23   & 1.11  & 1.23  & 4.7  & \: -0.06  & -     & 1.3   & 6.6     \\ \hline
  %
  D \:  & 10  & \: 72.5  & 117.5 & -0.24   &  1.09 &  4.95 & 18.8 & \: -0.157 & 116   & 1.3   & 7.4     \\ \hline
  %
  E  \: & 15  & \: 70    & 115   & -0.24   & 1.26  & 3.47  & 13.2 & \: -0.29  & 59    & 1.4   & 8.0     \\ \hline
  %
  %
  %
  F \:  & 20  & \: 67.5  & 112.5 & -0.25   & 1.19  & 0.642 & 2.4  & \: -0.337 & 47    & 1.2   & 8.0     \\ \hline
  %
  G \:  & 30  & \: 62.5  & 107.5 & -0.26   & 1.17  & 12.6  & 47.9 & \: -0.394 & 48    & 1.2   & 8.2     \\ \hline
  \hline

 \end{tabular}%
 \label{samples}%
\end{table}
\end{center}
\end{widetext}

It generates pulses with a spectral linewidth of about 3-4~meV and a duration of about 600~fs. 
The central wavelength of the laser  is tuned to excite electron-hole pairs at the Fermi energy of 
the two-dimensional electron gas in the samples.
The pulse train from the laser system is split into the pump beam, which is circularly polarized,
and the weaker,  linearly polarized probe beam. Both beams are focused onto the sample
surface at near-normal incidence by an achromat, and the Kerr rotation angle of the reflected
probe beam is measured using an optical bridge detector. The typical power density used in
the experiments was about 40~W/cm$^2$ for the pump beam, and 4~W/cm$^2$ for the probe beam.
A lock-in scheme is used to increase the sensitivity of the experiment. The samples were mounted in the $^3$He insert of an
optical cryostat and cooled by $^3$He gas to a nominal temperature of 4.5~K. Magnetic fields of up to 10~Tesla were applied in the
sample plane during the TRKR measurements.

For photoluminescence  measurements, the
Ti-Sapphire laser operated in cw mode and was tuned to higher energies to excite the samples nonresonantly.
The excitation density for PL measurements was about 4~W/cm$^2$. The PL emitted
from the samples was collected using a grating spectrometer and a charge-coupled device (CCD)
sensor. All PL measurements were performed without applied magnetic fields at a nominal
sample temperature of 4.5~K.
For PLE measurements, a tunable cw Ti-Sapphire laser system was used.
The PL intensity of the low-energetic flank of the sample was recorded
using the spectrometer and the CCD sensor as a function of the laser
energy of the Ti-Sapphire laser.

\subsection{Combined PL and PL excitation experiments}

\begin{figure}[t]
\includegraphics[width=0.45\textwidth]{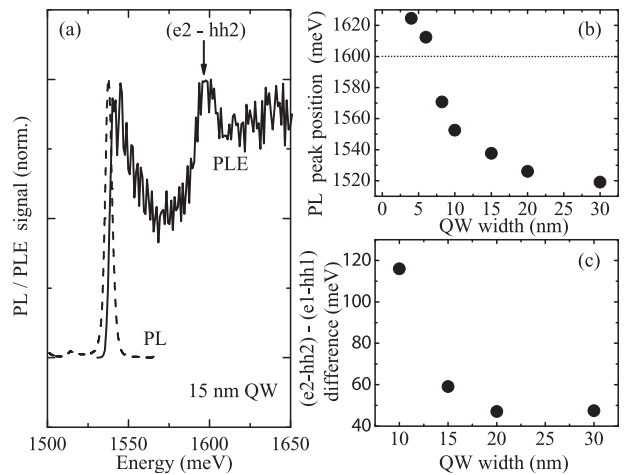}
\caption{(a) PL (black dashed line) and PLE (red solid line) spectra of
sample E (15~nm wide QW). The arrow indicates the transition
energy of the (e2-hh2) transition observed in the PLE spectrum. (b) PL
peak energy as a function of the QW width. The dotted line indicates the
transition energy for which the electron $g^*$ factor changes its
sign.~\cite{yugova} (c) Energy difference between (e1-hh1) and
(e2-hh2) transitions as a function of the QW width.}
\label{PL}
\end{figure}

To determine the electron confinement energies in the QWs, we performed
combined PL and PLE measurements on the sample series.
Figure~\ref{PL}(a) shows  PL and PLE spectra measured on
sample E
(15~nm wide QW). The PL spectrum shows a single
peak corresponding to the (e1-hh1) transition in the QW. In the
PLE spectrum, the onset of absorption at the Fermi energy of the
two-dimensional electron system in the QW is clearly visible
slightly above this PL peak. The pronounced maximum in the PLE
spectrum, indicated by the arrow, corresponds to the  (e2-hh2)
transition. Due to the increasing confinement in the narrower QWs,
the PL peak position shifts to higher energies, as
Fig.~\ref{PL}(b) demonstrates.  The energy difference between
(e1-hh1) and (e2-hh2) transitions as a function of the QW width, shown
in Fig.~\ref{PL}(c), and listed in Table~\ref{samples}, is
extracted from the PL and PLE measurements. It also increases as
the QW width is reduced. For QWs, which are thinner than
10~nm, the (e2-hh2) transition lies outside of the tuning range of
the Ti-Sapphire laser.

\subsection{Time-resolved Kerr rotation experiments}

TRKR measurements are used to determine the electron $g^*$ factor in our sample series.
It is well-established~\cite{Snelling91} that in GaAs/AlGaAs QWs  the electron $g^*$ factor depends on
the QW width, and even changes its sign from negative to positive values as a function of
the QW thickness. Two effects contribute to this dependence: The conduction band of GaAs is
nonparabolic, and the $g^*$ factor of an electron depends on its energy relative to
the conduction band edge.~\cite{yugova} Additionally, in narrow QWs, the
electron wave function has a sizeable amplitude within the AlGaAs barriers, leading to an
effective admixture of the (positive) $g^*$ factor in the barrier material to that (negative) of the
electron confined in the GaAs QW. In our TRKR measurements, we clearly observe this width
dependence [see Fig.~\ref{Kerr_gFactor}(a)]. 
The TRKR traces for different samples, measured at the same
magnetic field of 4~T, show the damped Larmor precession of the
optically oriented electron spin polarization about the applied
magnetic field. The precession frequencies are markedly different:
\begin{figure}[t]
\includegraphics[width=0.45\textwidth]{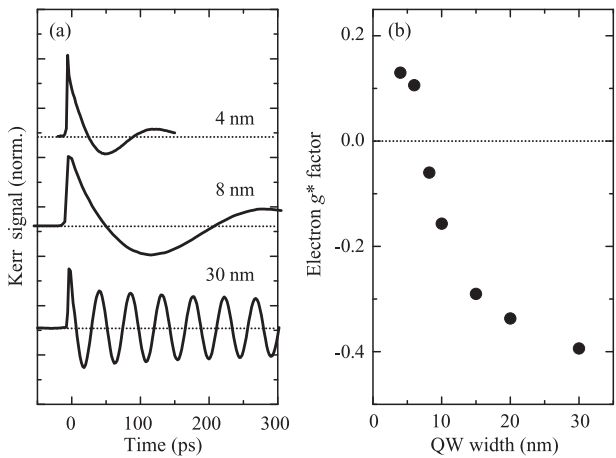}
\caption{(a) TRKR traces measured on 3 different samples 
($L_{\rm QW} = $ 4, 8 and 30~nm) with an applied 
in-plane magnetic field of 4~Tesla. (b) Electron $g^*$ factor as a function 
of the QW width extracted from the TRKR data. The sign of the $g^*$ factor, 
which cannot be directly extracted from the TRKR traces, has been 
inferred from the PL transition energies.}\label{Kerr_gFactor}
\end{figure}
For the widest sample, the precession frequency is high, it
decreases for narrower QWs, then increases again for the  most
narrow sample. The Larmor precession frequencies were determined
for all samples by measuring TRKR traces in magnetic fields
between 1 and 10~Tesla, and the amplitude of the $g^*$ factor was
calculated using a linear fit to the dispersion.
Figure~\ref{Kerr_gFactor}(b) gives the QW width dependence of the
$g^*$ factors in our samples. From TRKR measurements, the sign of
the $g^*$ factor is not immediately apparent,  although it can be
determined using slight modifications of a typical TRKR setup and
rigorous data analysis.~\cite{yang:152109} Here, we infer the sign
of the $g^*$ factor
from the PL transition energies observed for our samples. In their studies 
on the $g^*$ factor's dependence on
the QW width, Yugova et al.~\cite{yugova} observed the zero crossing of the $g^*$ factor for a
fundamental transition energy (e1-hh1) of 1600~meV (indicated by the dotted line in Fig.~\ref{PL}(b)).
In sample  C,
we observe the PL peak at 1572~meV, well below the value for zero crossing, and in
the thinner sample B
the PL peak is at 1616~meV, well above the value for zero crossing.
Hence, we assign negative $g^*$ factor values to the samples with nominal QW widths
between 30~nm and 8.2~nm, and positive $g^*$ factors to the two thinnest samples.
The TRKR measurements also allow us to  study the dependence of
the spin dephasing time (SDT), $T_2^*$, on the QW width.
Figure~\ref{TRKR_SDT_Omega}(a) shows a series of TRKR traces
measured on different samples without applied magnetic fields.
No simple correlation between the QW width and the SDT is
apparent,  as evidenced by the SDT data given in
Fig.~\ref{TRKR_SDT_Omega}(b) (note the logarithmic scale for the
SDT). In all samples, the dominant spin dephasing mechanism at low
temperatures is the D'yakonov-Perel mechanism~\cite{dp}, and in the
motional narrowing regime, the SDT is given by
\begin{equation}\label{DP_motional}
 \frac{1}{T_2^*} = \Omega^2 \tau_p\,\,\,,
\end{equation}
where $\tau_p$ is the momentum relaxation time, and
$\Omega$ is the precession
frequency due to the effective spin-orbit fields. As the QW width
is reduced, the magnitude of the Dresselhaus field increases due
to the momentum quantization along the growth axis, and the
magnitude of the Rashba field was also shown to increase
monotonously with decreasing QW width in the thickness range
investigated here.~\cite{PhysRevB.55.16293} The rising amplitude
of both spin-orbit fields should lead to a more rapid dephasing in
thinner QWs according to Eq.~(\ref{DP_motional}). We can
explain the non monotonous dependence of the SDT on the QW width
by taking into account the different momentum relaxation times in
our samples. The momentum relaxation times $\tau_p$, 
calculated from the mobility data measured at 4.2~K, are
listed in Table~\ref{samples}. Using these values, we calculate
$\Omega^2$ using Eq.~(\ref{DP_motional}). The results,
shown in Fig.~\ref{TRKR_SDT_Omega}(c), demonstrate a
near-monotonous increase of $\Omega^2$ with decreasing QW
width, as expected from the QW width dependence of the effective
spin-orbit fields.

\begin{figure}[t]
\includegraphics[width=0.45\textwidth]{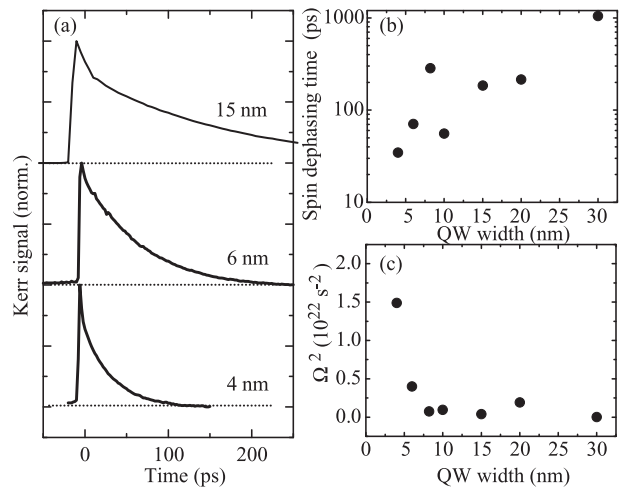}
\caption{(a) Zero-field TRKR traces measured on three different
samples with $L_{\rm QW} = $ 4, 6 and 15~nm. (b) Zero-field SDT as a function of the QW width (log.
scale). (c) $\Omega^2$ as a function of the QW
width.}\label{TRKR_SDT_Omega}
\end{figure}

\section{Linear MPGE}
\label{LMPGE}

The linear MPGE excited by linearly polarized radiation has been 
detected in all samples.
Under THz irradiation we observed a photocurrent signal $J_x$, which is
linearly increasing with rising magnetic field strength $B_y$ and changes 
its sign upon the
inversion of the magnetic field direction from ${B_y} > 0$ to ${B_y} < 0$, demonstrating the typical MPGE behavior.
\begin{figure}[t]
\includegraphics[width=0.5\textwidth]{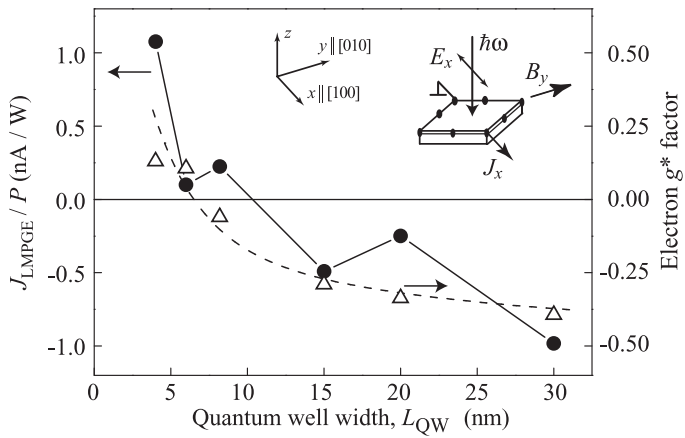}
\caption{Dependence of the linear MPGE (circles) on $L_{\rm QW}$ obtained at room temperature, $B_y = \pm 1$~T and photon energy $\hbar \omega = 4.4$~meV and corresponding $g^*$ factors (triangles, by TRKR).
The inset shows the experimental geometry for the LMPGE.}\label{fig2}
\end{figure}
As our experiments here are focused on the linear MPGE, 
we eliminate any possible background in our experiments by
\begin{equation}
\label{MPGE}
\bm{J}^{\rm L} = [\bm{J}(B_y) - \bm{J}(-B_y)]/2.
\end{equation}
Figure~\ref{fig2} shows  the photocurrent $\bm{J}^{\rm L}$ as a
function of $L_{\rm QW}$ obtained for magnetic fields of $\pm 1$~T.
For comparison the effective Land\'{e} factor $g^*$,  extracted from the  time-resolved Kerr rotation,
is also plotted (details of these measurements are already discussed in Sec.~\ref{kerr}).
As an important result Fig.~\ref{fig2} demonstrates that the SIA-induced photocurrent $J_x$,
similarly to the $g^*$ factor, changes its sign upon the variation of $L_{\rm QW}$.~\cite{footnote1}
However, there is a difference in the zero points: While the $g^*$ factor equals 
to zero at $L_{\rm QW} \sim 6.5$~nm,
the current vanishes for $L_{\rm QW} \sim  10$~nm.
It will be shown below that this shift between the inversion points
as well as the MPGE's sign inversion can be well described by
the interplay of spin and orbital mechanisms in the current formation.

\begin{figure}[t]
\includegraphics[width=0.38\textwidth]{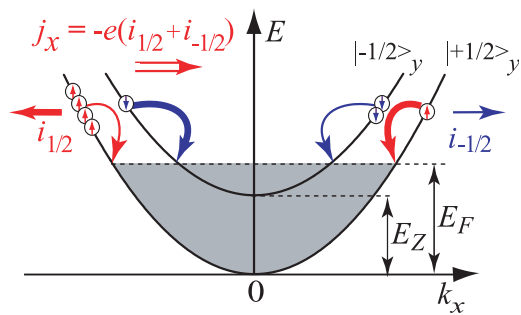}
\caption{Spin based model for the linear MPGE. See details in the text.}\label{fig5a}
\end{figure}

To explain qualitatively our results we 
describe the basic physics of these mechanisms.
We start with the spin-related mechanism. The generation of 
a spin-polarized current due to the MPGE may
be discussed in the frame of a recently proposed model for the
spin-dependent asymmetric energy relaxation of a non-equilibrium
electron gas heated by, e.g., THz  or microwave
radiation.~\cite{naturephysics06,APL2010} Free electrons are
excited to higher energy states by absorbing radiation and then relax into an equilibrium
state by emitting phonons. Figure~\ref{fig5a} sketches the hot
electron energy relaxation processes in the two spin subbands
($s_y =\pm 1/2)$ that are split due to the Zeeman effect in the
presence of an external magnetic field.
The electron relaxation from higher to lower energies is shown by bent arrows.
In gyrotropic media, like GaAs low-dimensional semiconductors, the spin-orbit 
interaction makes the scattering probability 
spin-dependent:~\cite{naturephysics06}
\begin{equation}
\label{k_sigma}
    W_{\bm k' \bm k} = W_0 \left\{1+ \xi \, [\bm \sigma \times (\bm k+\bm k')]_z \right\},
\end{equation}
where $W_0$ is the symmetric part of the scattering probability, which
determines the mobility (we consider SIA only). Here $\bm{k}$ and $\bm{k}^\prime$ are the
initial and scattered electron wave vectors and $\bm{\sigma}$ is
the vector composed of the Pauli matrices. Thus, the electron
transitions to positive and negative $k_x^\prime $-states occur
with different probabilities causing an imbalance in the
distribution of carriers in both subbands between positive and
negative $k_x$-states. This is shown by different thicknesses of
the bent arrows in Fig.~\ref{fig5a}. At zero magnetic, field the
asymmetry of the electron-phonon interaction would lead to a spin
current, but not an electric current: This is due to the fact that
the oppositely directed electron fluxes, $i_{\pm 1/2}  \propto \xi$, are of
equal strength and therefore, compensate each other. The presence
of an in-plane magnetic field $B_y$, however, leads to the Zeeman
splitting of the $s_y =\pm 1/2$ subbands. As a consequence, the
electron densities $n_{+1/2}$ and $n_{-1/2}$ in both spin subbands
become different, and the fluxes $i_{\pm 1/2}$ do not longer
compensate each other and therefore a net electric current
results. Obviously, this current, classified as linear MPGE, is
spin-polarized and its value is proportional to the Zeeman spin
splitting induced by the magnetic field. 
Its QW width dependence is described by 
the product $g^* \, \xi$.

In order to estimate the parameter $\xi$ caused by SIA, we take into account that 
the remote impurities create an electric field $\cal E$ along the growth direction, 
leading to the asymmetry of the QW. Therefore, the eigenstates in this
structure are superpositions of the states of the rectangular QW.
As a result, the envelope wave functions in the first and second
subbands of size quantization $\phi_{1,2}(z)$ are given by
\begin{eqnarray}
\label{E_mixing}
  \phi_{1}(z) &=& \varphi_1(z)+{e{\cal E}z_{21} \over E_{21}}\varphi_2(z),\\
  \phi_{2}(z) &=& \varphi_2(z)-{e{\cal E}z_{21} \over E_{21}}\varphi_1(z). \nonumber
\end{eqnarray}
Here $z$ is the growth direction, $\varphi_{1,2}(z)$ are the functions 
of size quantization in the ground and the first  excited subbands of the rectangular QW
of width $L_{\rm QW}$, $E_{21}$ is the energy separation between
these subbands, and $z_{21}$ is the coordinate matrix element
calculated between these states. 
This leads to $\xi(L_{\rm QW}) \sim z_{21} / E_{21}$, and
to the following dependence of this  contribution to the linear MPGE current:
\begin{equation}
\label{linspin}
 \bm{j}^{\rm L}_{\rm spin}(L_{\rm QW}) \sim g^*(L_{\rm QW}) \, {z_{21}\over E_{21}}\,\,\,.
\end{equation}
Here we  disregard a weak dependence of the scattering probability on the QW width.

In a set of samples with similar structure inversion asymmetry, like in our samples,
where SIA is controlled by the asymmetric doping, the MPGE should
vanish for the sample with zero $g^*$ factor and change its sign
upon a QW width variation. 
Thus, the spin-related mechanism of the
MPGE describes well the appearance of the sign inversion. However,
it can not explain the fact that the MPGE's sign inversion takes
place for an about 4~nm broader QW than the inversion of the 
$g^*$ factor.~\cite{footnote2}
The explanation of this fact
requires an additional spin-independent orbital contribution to
the MPGE, recently addressed by Tarasenko.~\cite{Tarasenko_orbital}

\begin{figure}[t]
\includegraphics[width=0.4\textwidth]{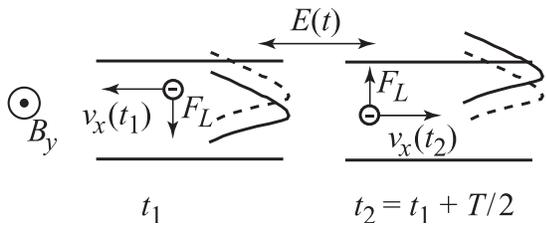}
\caption{Orbital model for the linear MPGE. Solid and dashed 
lines show the electron wavefunction with and without radiation, respectively.}\label{fig5b}
\end{figure}

A process resulting in the orbital linear MPGE is illustrated in Fig.~\ref{fig5b}.
Here, we consider the in-plane motion of carriers subjected to the
high-frequency electric field (e.g., THz radiation) applied along
the $x$ direction and an in-plane magnetic field along the
$y$ direction. For linearly polarized radiation, the electric
field $\bm{E}(t)$ leads to a back and forth motion of electrons
with velocity $\bm{v}(t)$ as illustrated in Fig.~\ref{fig5b}
for two different times.  Without external field the electron's
wave function is localized closer to one
interface due to the asymmetrical doping, and its momentum relaxation rate $1/\tau_p$ 
has a contribution controlled by scattering by this interface.
The orbital MPGE originates from the combined action of
electric and magnetic fields, described in the following. If
$\bm{E}(t)$ is applied, at a certain time $t_{1}$, the electron is
accelerated by the in-plane  $ac$ electric field along the
negative $x$ direction. At the same time, the electron with
velocity ${v_x}$ is subjected to the magnetic field ${B_y}$.  This
results in the Lorentz force $\bm F_{\rm L} =  e  (\bm v \times \bm B)$
which shifts the wave function to the center of the quantum well
and thus decreases the scattering probability (see
Fig.~\ref{fig5b}). Half a period later, at $t_2 = t_1 + T/2$,  the
electron velocity gets reversed so that the direction of the
Lorentz force reverses as well. Now, the electron wave function is
pushed closer to the interface, which increases the scattering
rate and decreases the momentum relaxation. The resulting
imbalance of the relaxation times for the motion along positive
and negative $x$ direction causes a net electric current
proportional to the magnetic field strength. Obviously this
mechanism yields a photocurrent, which does not change its sign with a variation of the QW width.
Microscopically, the orbital contribution to the linear MPGE current 
$\bm j^{\rm L}_{\rm orb}$ is caused by  the effect of a magnetic field on the
scattering. The in-plane magnetic field
results in a $\bm k$-dependent mixing of the eigenstates $\phi_1$
and $\phi_2$, and the envelope in the ground subband  becomes
$k_\alpha B_\beta$-dependent~\cite{Tarasenko_orbital}
\begin{equation}
\label{kB_mixing}
\psi_{1\bm k}(z)=\phi_{1}(z) + (\bm k \times \bm B)_z {e\hbar\over m} {z_{21} \over E_{21}}\phi_{2}(z).
\end{equation}
This leads to the following contribution in the scattering probability
\begin{equation}
\label{kB_scattering}
    W_{\bm k' \bm k} = W_0 \left\{1+ \zeta \, [\bm B \times (\bm k+\bm k')]_z \right\},
\end{equation}
where the parameter $\zeta$ is caused by SIA.
The dependence $\zeta(L_{\rm QW})$ determines the dependence of $\bm{j}^{\rm L}_{\rm orb}$ on the quantum well width yielding
\begin{equation}
\label{LMPGE_orb}
    \bm{j}^{\rm L}_{\rm orb}(L_{\rm
QW})  \sim \left({z_{21} \over E_{21}}\right)^2.
\end{equation}
The quadratic dependence on the parameter $z_{21} / E_{21}$ appears because, 
both, the electric field of the impurities and the magnetic field, lead to a mixing 
of the ground and first excited levels, Eqs.~\eqref{E_mixing} and~\eqref{kB_mixing}. 
Here, like for Eq.~\eqref{linspin}, we also assume that additional factors originating from the scattering asymmetry depend weakly on the QW width.

\begin{figure}[t]
\includegraphics[width=0.35\textwidth]{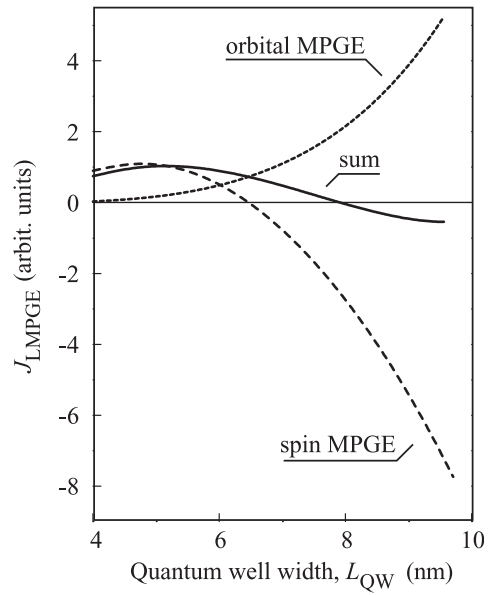}
\caption{Dependence of the spin-dependent (triangles) and orbital (squares) 
contributions to MPGE on QW width.}\label{fig:theory}
\end{figure}

On the phenomenological level both mechanisms are described by the same equations
and the total current is given by the sum of both contributions
\begin{equation}
\label{j}
    \bm j^{\rm L} = \bm j^{\rm L}_{\rm spin} + \bm j^{\rm L}_{\rm orb}.
\end{equation}
The phenomenological similarity hinders the decomposition of both terms, 
because the spin contribution $\bm j^{\rm L}_{\rm spin}$ 
and the orbital one $\bm j^{\rm L}_{\rm orb}$ behave identically under a variation of the 
radiation's polarization state and the orientation of the magnetic field relative to the crystallographic axes.
Our above consideration shows, however, that the behavior of the photocurrent upon a variation of the
QW width allows us to distinguish between these two basically different mechanisms.
The observed sign inversion clearly demonstrates that the
total current almost reflects the behavior of the $g^*$ factor and,
therefore, is in most samples dominated by the spin mechanism.
The small shift of the MPGE's inversion point to a larger $L_{\rm QW}$ compared to the 
one of the $g^*$ factor, however,
demonstrates that the orbital MPGE also contributes and its
magnitude is smaller but still comparable to that of the spin MPGE.
For a QW with $L_{\rm QW}$ of about 6.5~nm  the $g^*$ factor is equal 
to zero and the MPGE is solely caused by the orbital mechanism.

The dependence of $\bm{j}^{\rm L}_{\rm orb}$ on the quantum well width, calculated
after Eq.~\eqref{LMPGE_orb} is plotted in Fig.~\ref{fig:theory}.
Figure~\ref{fig:theory} shows that this current rises with increasing  $L_{\rm QW}$ due to the 
increase of the mixing parameter $z_{21} / E_{21}$ with $L_{\rm QW}$
but does not change its sign. 
This behavior is explained as follows: In narrow QWs the electric and magnetic fields can not efficiently mix size-quantized states because of their large energy separation $E_{21}$. 
In wider QWs the confinement is weaker, and the photocurrent increases. 
This strong dependence is valid for not too wide QWs, where only the ground 
subband is occupied. In very wide QWs, the photocurrent dependence on the QW width 
has a maximum and then the photocurrent tends to zero since 
in bulk GaAs photogalvanic effects are absent.

Figure~\ref{fig:theory} also shows the interplay of both contributions. 
To obtain this curve, we have normalized both contributions in such a way that the total current
vanishes for the QW width where the photocurrent's sign inversion has been detected.
One can see that the total photocurrent vanishes for a larger QW width  than
the $g^*$ factor. This is due to the orbital contribution.

\section{Circular MPGE}
\label{CMPGE}

\begin{figure}[t]
\includegraphics[width=0.45\textwidth]{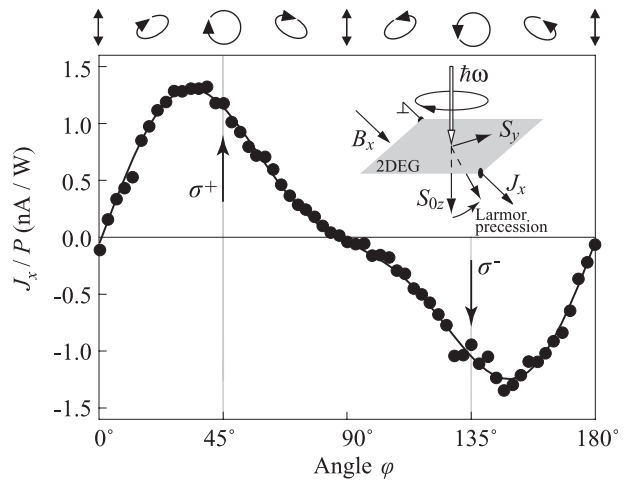}
\caption{Helicity dependence of the MPGE obtained for sample A at room temperature, $\left|\bm{B}\right| = 1$~T and a photon energy of $\hbar \omega = 4.4$~meV. The ellipses on top illustrate the polarization states for various angles of the $\lambda/4$ plate, $\varphi$. 
The inset depicts the spin-galvanic effect.}\label{fig3}
\end{figure}

\begin{figure}[t]
\includegraphics[width=0.5\textwidth]{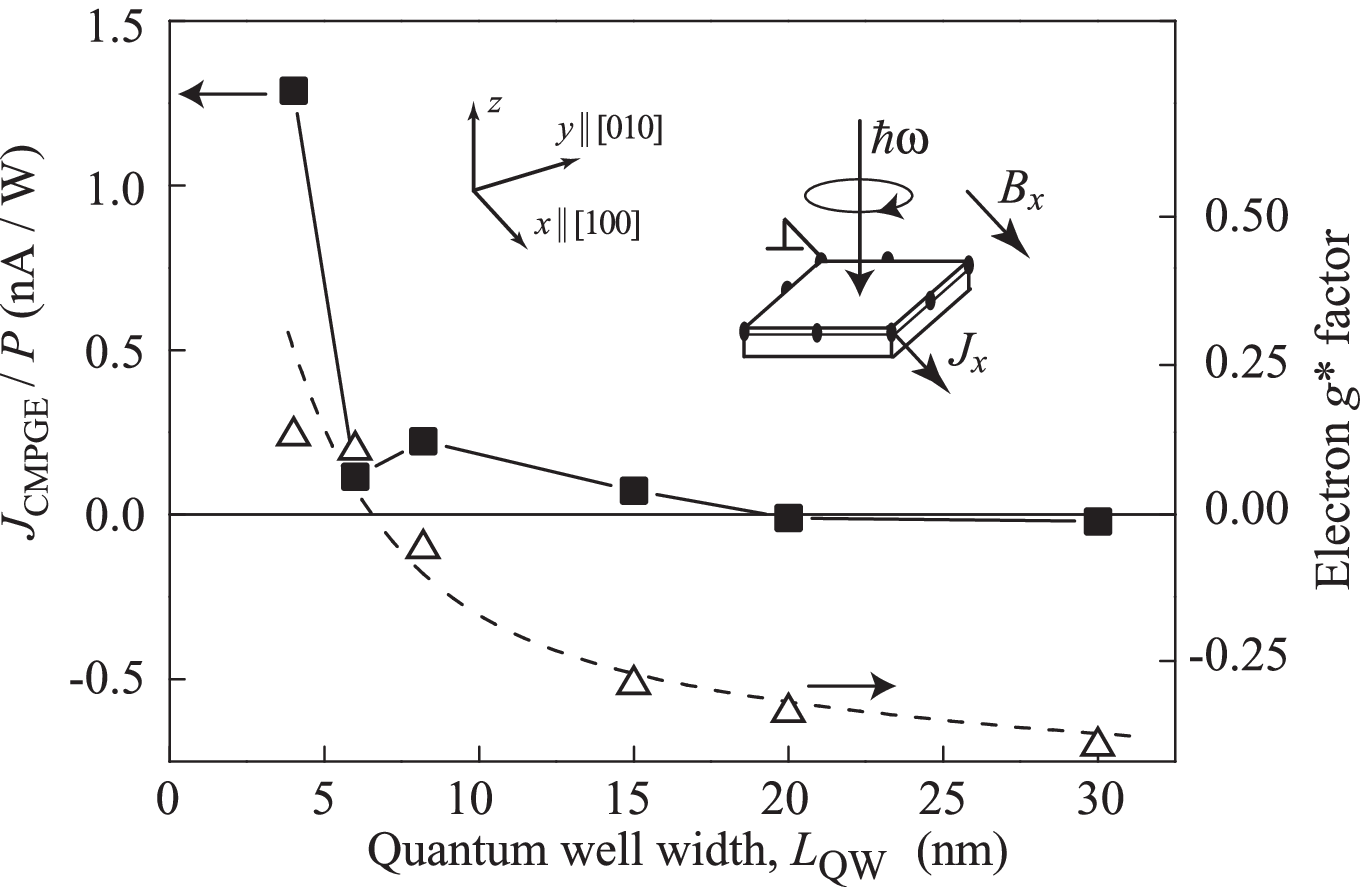}
\caption{Dependence of the circular MPGE (squares) on $L_{\rm QW}$ 
obtained at room temperature, $\left|\bm{B}\right| = 1$~T and a photon 
energy of $\hbar \omega = 4.4$~meV
and corresponding $g^*$ factors (triangles, by TRKR).
The inset shows the experimental geometry for the CMPGE.}\label{fig4}
\end{figure}

The sign inversion upon a variation of the QW width is also obtained for the
SIA-induced circular MPGE. 
Figure~\ref{fig3} shows the helicity dependence of the
photocurrent $J_x$, which clearly demonstrates the fingerprint of the circular MPGE: The sign inversion upon switching the radiation's helicity $P_{\rm circ}$ from $+1$ to $-1$ 
at $\varphi = 45^{\circ}$ ($\sigma^{+}$) and $\varphi =
135^{\circ}$ ($\sigma^{-}$), respectively.
We note, that the whole polarization dependence stems from the
interplay of the circular and linear MPGE. In agreement with the
phenomenological theory~\cite{MPGE2005} the signal can be well fitted by 
$J_x = A P_{\rm circ} + B  (1 + \cos 4\varphi)  + C  \sin 4\varphi$.
Here the first term, given by the parameter $A$, is due to the
circular MPGE  and the following two, proportional to the
coefficients $B$ and $C$, stem from the linear MPGE discussed above 
and vanish for  circularly polarized radiation.
Figure~\ref{fig4} shows  the  quantum well width dependence of the
circular photocurrent.
In order to extract the circular MPGE from the total current and
to ensure that no spurious currents contribute to the signal
we used the fact that the circular MPGE
reverses its sign upon switching the radiation's helicity
from $+1$ ($\sigma^+$) to $-1$ ($\sigma^-$) 
and obtain the circular photocurrent after
\begin{equation}
 J_x^{\rm C} = [J_x(\sigma^+) -  J_x(\sigma^-)] / 2,
\label{equCMPGE}
\end{equation}
where $J_x(\sigma^+)$ and  $J_x(\sigma^-)$ are photocurrents measured at $\sigma^+$- and $\sigma^-$-polarized excitation.
Because the circular MPGE also reverses its sign upon 
the inversion of the magnetic field 
direction, the current after Eq.~(\ref{equCMPGE}) is further treated 
similarly to the linear MPGE according to Eq.~(\ref{MPGE}).
For the circular MPGE, we also observed that the photocurrent reverses its sign
at a certain QW width.~\cite{footnote3} Similarly to the behavior of the linear
MPGE, the sign inversion of the photocurrent in Fig.~\ref{fig4} does not coincide with
that of the Land\'{e} factor and takes place at $L_{\rm QW}\sim
15$~nm. We show below that, similarly to the linear MPGE discussed earlier, 
this fact indicates the interplay of the spin and orbital 
mechanisms in the circular MPGE.

The spin mechanism of the circular MPGE is based on the
spin-galvanic effect.~\cite{Nature02} This effect results in the generation 
of an electric current due to a non-equilibrium spin polarization and 
is caused by its asymmetric spin relaxation.
The spin-galvanic current and the average non-equilibrium spin {$\bm S$} are
related by a second rank pseudotensor, with components
proportional to the parameters of the spin-orbit splitting, as follows
\begin{equation} 
j^{\rm C}_{\rm spin,\alpha} =
\sum_{\gamma}Q_{\alpha \gamma} S_{\gamma}
\label{equ22}.
\end{equation}
For (001)-grown zinc-blende structure based  QWs of C$_{2v}$-symmetry, the
spin-galvanic effect is allowed only for a spin polarized electron gas with
spins oriented in the QW plane. In the coordinate system parallel to the cubic axes, 
Eq.~(\ref{equ22}) reduces to $j_{x}^{\rm SIA} = Q_{xy} S_y$ and $j_{y}^{\rm BIA} = Q_{yy}S_y$,
with $j_{x}^{\rm SIA}$ and $j_{y}^{\rm BIA}$ proportional to the Rashba 
and the Dresselhaus constants, respectively.~\cite{BIASIAPRB}
The spin-galvanic effect generally does not require optical excitation.
In fact, the non-equilibrium spin
can be achieved by both optical and non-optical methods, e.g., by electrical 
spin injection.

The spin-dependent contribution $\bm{j}^{\rm C}_{\rm spin}$ 
stems from the asymmetry in the spin relaxation, arising due to 
the Rashba spin-orbit splitting of the electronic ground subband 
and being linear in the in-plane wave vector $\bm k$:
\begin{equation}
    E_{1\uparrow}-E_{1\downarrow} = \hbar\Omega\,\,\,.
\end{equation}
Here the SIA-induced $\bm k$-linear spin-orbit splitting $\hbar\Omega$ is measured by TRKR technique, Sec.~\ref{kerr}. 

An optical method, which provides a non-equal population of an uniform 
distribution in both spin subbands was proposed in Ref.~\onlinecite{Nature02}. 
This method, sketched in the inset of
Fig.~\ref{fig3}, is based on the optical excitation with
circularly polarized light yielding a steady-state spin
orientation $S_{0z}$ in the growth direction. The in-plane magnetic field 
then rotates, due to the Larmor precession, the optically oriented spins 
into the plane of the 2DES. If $\bm B$ is oriented along the $x$ axis we 
obtain a non-equilibrium spin polarization $S_y$ which reads after time
averaging~\cite{Nature02}
\begin{equation}
S_y = -\frac{\omega_L\tau_{s \perp}}{1 + (\omega_L \tau_{s})^2}\:S_{0z}\:,
\end{equation}
where $\tau_s = \sqrt{\tau_{s\parallel} \tau_{s\perp} }$ and
$\tau_{s \parallel}, \tau_{s \perp}$ are the longitudinal and
transverse electron spin relax\-ation times, 
the Larmor frequency is given by $\omega_L = g^*\mu_{\rm B} B_x / \hbar$, 
$\mu_{\rm B}$ is the
Bohr magneton, and $S_{0z} = \tau_{s\parallel}\dot{S}_z$ is the
steady state electron spin polarization in the absence of a
magnetic field. According to Eq.~(\ref{equ22}) the in-plane spin
polarization $S_y$ causes a net electric current 
in the direction along the magnetic field. 
For the investigated QWs at room temperature $\omega_L\tau_s \ll 1$ and, therefore, 
the photocurrent is proportional to $g^* B_y$.

The QW width dependence of $\bm{j}^{\rm C}_{\rm spin}$ is given by
\begin{equation}
\label{spincirc}
    \bm{j}^{\rm C}_{\rm spin}(L_{\rm QW}) \sim g^*(L_{\rm QW}) \, \Omega(L_{\rm QW})\, \tau_s^2(L_{\rm QW}).
\end{equation}
Equation~\eqref{spincirc} shows that this contribution is proportional to the 
$g^*$ factor and, consequently, reflects its behavior upon a QW 
width variation.~\cite{footnote5} Despite the QW width dependence 
of $\bm{j}^{\rm C}_{\rm spin}$ does not coincide with $g^*(L_{\rm QW})$, this contribution 
to the photocurrent vanishes in QWs with $g^* = 0$.
The fact that in GaAs/AlGaAs QWs the current, similarly to the $g^*$ factor,
changes its sign upon a variation of $L_{\rm QW}$ qualitatively explains 
the MPGE's sign inversion
but can not clarify the difference in the two zero points.
To explain this shift we should again address the possible
contribution of the orbital MPGE.

An orbital mechanism yielding a helicity driven MPGE current was recently 
suggested in Ref.~\onlinecite{Tarasenko_orbitalMPGE2}
and is given by
\begin{equation}
    j^{\rm C}_{\rm orb, \alpha} = P_{\rm circ} |E_0|^2 \sum_\gamma R_{\alpha\gamma} B_\gamma.
\end{equation}
The second rank pseudotensor $\bm R$ has the same space symmetry
properties as  the pseudotensor $\bm Q$, describing the
spin-galvanic effect. However, the tensor $\bm R$ is invariant
under time inversion. Microscopically, the orbital contribution to
the circular MPGE appears similarly to the one to the linear
MPGE current described above. The current is caused by the action
of the Lorentz force on the orbital motion of the 2D electrons in
the radiation field. Under irradiation with circularly polarized
light electrons perform a cyclic motion and, due to SIA,
the presence of an in-plane magnetic field 
forces them  to flow predominantly along the direction of $\bm B$. 
Reversing the radiation's helicity, changes the current direction. The microscopic
theory of this effect is given in Ref.~\onlinecite{Tarasenko_orbitalMPGE2}. The resulting
current 
$\bm j^{\rm C}_{\rm orb}$,  
is caused by the $\bm B$-dependent corrections to the scattering
probability, Eq.~\eqref{kB_scattering}, as well as the
corresponding contribution to the  linear MPGE.
The photocurrents due to the circular and linear MPGE are linked to each other according to
$\bm{j}^{\rm C}_{\rm orb} \sim {\bm{j}^{\rm L}_{\rm orb} \, \omega\tau_p}$ at $\omega\tau_p \ll 1$ (Ref.~\onlinecite{Tarasenko_orbital}). 
Using Eq.~\eqref{LMPGE_orb}, we can, consequently, estimate the QW width 
dependence of $\bm{j}^{\rm C}_{\rm orb}$ as
\begin{equation}
    \bm{j}^{\rm C}_{\rm orb}(L_{\rm QW}) \sim \left({z_{21} \over E_{21}}\right)^2.
\end{equation}
This equation shows that the orbital contribution to the circular MPGE
has a constant-sign dependence on the QW width. 
Thus, the observed sign inversion of the circular MPGE and the shift between 
its zero crossing and that of the $g^*$ factor agrees with the picture of the interplay 
of comparable spin and orbital contributions to this phenomenon. 

Both mechanisms of the circular MPGE contribute to the total current $\bm j^{\rm C} =
\bm{j}^{\rm C}_{\rm spin} + \bm j^{\rm C}_{\rm orb}$  and are described phenomenologically
by similar equations. The observed sign inversion proves
that, in correspondence with the results of the linear MPGE, the
dominant contribution comes from the spin-galvanic effect. 
While the spin-galvanic effect dominates the current for most of the investigated 
samples, the existence of the orbital circular MPGE is also clearly demonstrated. In
particular, for QWs with a width of about 6.5~nm the spin-galvanic
effect vanishes and the current is caused solely by the orbital circular MPGE.

\section{Summary}
\label{summary}

To summarize, our experiments clearly demonstrate that, both, linear and circular
MPGE in GaAs/AlGaAs QW structures result from spin and orbital 
contributions. 
Our experiments show that  for most quantum well 
widths the MPGE is mainly driven by spin-related mechanisms, which 
result in a photocurrent 
proportional to the $g^*$ factor. For structures with a vanishingly 
small $g^*$ factor, however, the MPGE caused by 
orbital mechanisms is clearly observed.
Our work demonstrates that a variation of the electron $g^*$ factor
by different means, like varying the QW width, as it is done here, 
doping with magnetic impurities or 
using narrow band materials, where $g^*$ and the spin-orbit interaction are enhanced, can 
be used for the separation of these qualitatively different mechanisms. Moreover, the
orbital MPGE can be studied independently in materials with a vanishingly small
spin-orbit interaction, like Si-based metal-oxide-semiconductor low dimensional structures.

We thank  S.A.~Tarasenko and E.L. Ivchenko for fruitful discussions and support, as well as M.~Schmalzbauer. 
This work is supported by the DFG via programs SPP~1285, SFB~689, 
by the Linkage Grant of IB of BMBF at DLR, 
Russian Ministry of Education and Sciences, 
RFBR and ``Dynasty'' foundation --- ICFPM.


\begin{thebibliography}{99}


\bibitem{LL5} L. D. Landau and E. M. Lifshits, Course of theoretical physics V ``Statistical Physics''.

\bibitem{Exc} \textit{Excitons}, edited by E. I. Rashba and M. D. Sturge (North-Holland, Amsterdam, 1982).

\bibitem{Fabian08} J. Fabian, A. Matos-Abiague, C. Ertler, P. Stano, and I. Zutic,
Acta Physica Slovaca \textbf{57}, 565 (2007), arXiv:cond-mat/0711.1461.

\bibitem{Ivchenkobook2} E.L.~Ivchenko, \textit{ Optical
Spectroscopy of Semiconductor Nanostructures} (Alpha Science Int.,
Harrow, UK, 2005).

\bibitem{GanichevPrettl} S.~D.~Ganichev and W.~Prettl,
\textit{Intense Terahertz Excitation of Semiconductors} (Oxford
University Press, Oxford 2006).

\bibitem{Winkler06} R. Winkler,
\textit{Spin-Dependent Transport of Carriers
in Semiconductors}, in  \textit{Handbook of Magnetism and Advanced
Magnetic Materials} (John Wiley \& Sons, NY 2007); ibido arXiv cond-mat: 0605390 (2006).

\bibitem{Ivchenko08} E. L. Ivchenko and S.~D.~Ganichev, 
\textit{ Spin Photogalvanics} in \textit{Spin Physics in Semiconductors}, 
edited by M.~I. D'yakonov (Springer, Berlin 2008).

\bibitem{SSTreview} V. V. Bel'kov  and S. D.~Ganichev,
Semicond. Sci. Technol. \textbf{23}, 114003 (2008). 

\bibitem{Handbook} V. V. Bel'kov and S. D.~Ganichev,
\textit{Zero-bias spin separation}, in \textit{Handbook of
Spintronic Semiconductors}, edited by W.~M.~Chen and I.~A.~Buyanova
 (Pan Stanford Publishing, Singapore 2010).

\bibitem{naturephysics06} S.~D.~Ganichev, V.~V.~Bel'kov, S.~A.~Tarasenko, S.~N.~Danilov, S.~Giglberger, C.~Hoffmann, E.~L.~Ivchenko, D.~Weiss, W.~Wegscheider, C.~Gerl, D.~Schuh, J.~Stahl, J.~de~Boek, G.~Borghs and W.~Prettl,
{Nature Physics} \textbf{2}, 609 (2006).

\bibitem{condmat110} V.~V.~Bel'kov, P.~Olbrich, S.~A.~Tarasenko, D.~Schuh, W.~Wegscheider,
T.~Korn, C.~Sch\"uller, D.~Weiss, W.~Prettl, and S.~D.~Ganichev, {Phys. Rev. Lett.}    {\bf 100}, 176806 (2008).

\bibitem{Nature02} S.~D.~Ganichev, E.~L.~Ivchenko, V.~V.~Bel'kov, S.~A.~Tarasenko, M.~Sollinger, D.~Weiss, W.~Wegscheider, and W.~Prettl,
Nature {\bf 417}, 153 (2002).

\bibitem{SGEopt}
S.~D.~Ganichev, Petra~Schneider, V.~V.~Bel'kov, E.~L.~Ivchenko,
S.~A.~Tarasenko, W.~Wegscheider, D.~Weiss, D.~Schuh, B.~N.~Murdin,
P.~J.~Phillips,  C.~R.~Pidgeon,
D.~G.~Clarke, M.~Merrick,  P.~Murzyn, 
 E.~V.~Beregulin, and W.~Prettl,
Phys.~Rev.~B. 
{\bf 68}, 081302 (2003).

\bibitem{review2003spin} S.~D.~Ganichev  and W.~Prettl,
J. Phys.: Condens. Matter {\bf 15}, R935 (2003).

\bibitem{Tarasenko_orbital} S.~A.~Tarasenko,
Phys. Rev. B \textbf{77}, 085328 (2008).

\bibitem{Tarasenko_orbitalMPGE2} S.~A.~Tarasenko,
arXiv cond-mat: 1009.0681v1 (2010).


\bibitem{Dyakonovbook} \textit{Spin Physics in Semiconductors}, edited by M.~I.~Dyakonov 
(Springer, Berlin 2008).

\bibitem{Awschalombook2010} \textit{Semiconductor Spintronics and Quantum Computation }
eds. D.~D. Awschalom, D. Loss, and N. Samarth  (Springer, Berlin 2010).

\bibitem{Wu2010} M. W. Wu, J. H. Jiang, and M. Q. Weng,
Phys. Reports \textbf{493},  61 (2010).

\bibitem{Snelling91} M. J. Snelling, G. P. Flinn, A.~S. Plaut, R.~T. Harley, A.~C. Tropper, R. Eccleston, 
and C.~C. Phillips,
Phys. Rev. B \textbf{44}, 11345 (1991).

\bibitem{Ivchenko_g_factor}  E.~L.~Ivchenko, A. A. Kiselev, and M. Willander,
Sol. St. Com. \textbf{102}, 375 (1997).

\bibitem{yugova} I. A. Yugova, A. Greilich, D.~R. Yakovlev, A.~A. Kiselev, M. Bayer, V.~V. Petrov, Yu.~K. Dolgikh, D. Reuter, and A.~D. Wieck,
Phys. Rev. B \textbf{75},
245302 (2007).

\bibitem{JETP1982} S.~D.~Ganichev, S.~A.~Emel'yanov, and I.~D.~Yaroshetskii,
Sov. Phys. JETP Lett. {\bf  35},  368 (1982).

\bibitem{BIASIAPRB} S.~Giglberger, L.~E.~Golub,  V.~V.~Bel'kov, S.~N.~Danilov, D.~Schuh, Ch.~Gerl,
F.~Rohlfing,  J.~Stahl, W.~Wegscheider, D.~Weiss,
W.~Prettl, and S.~D.~Ganichev,
Phys. Rev. B \textbf{75} 035327 (2007).

\bibitem{LechV12} 
V.~Lechner, L.~E.~Golub, P.~Olbrich, S.~Stachel,
D.~Schuh, W.~Wegscheider, V.~V.~Bel'kov, and S.~D.~Ganichev, 
Appl. Phys. Lett. {\bf 94}, 242109 (2009).

\bibitem{Harley07}W. J. H. Leyland, G. H. John, R. T. Harley, M.~M. Glazov, E.~L.
Ivchenko, D.~A. Ritchie, I. Farrer, A.~J. Shields, and M. Henini. 
Phys. Rev. B {\bf 75}, 165309 (2007).

\bibitem{yang:152109} C.~L.~Yang, J. Dai, W.~K. Ge, and X. Cui, Appl. Phys. Lett. \textbf{96}, 152109 (2010).

\bibitem{dp} M.~I.~Dyakonov and V.~I.~Perel, Sov. Phys. - JETP \textbf{33}, 1053
(1971).

\bibitem{PhysRevB.55.16293} E.~A.~de Andrada e Silva, G.~C.La~Rocca,  
and F. Bassani,  Phys. Rev. B \textbf{55}, 16293
(1997).

\bibitem{footnote1} Note, that $\chi$ of the sample with $L_{\rm QW} = 4$~nm is twice as large as in the other samples, therefore SIA and consequently the SIA-induced current is enhanced compared to the samples with larger $L_{\rm QW}$.

\bibitem{APL2010} C. Drexler, V.~V. Bel'kov, B. Ashkinadze, P. Olbrich, C. Zoth,
V. Lechner, Ya.~V. Terent'ev, D.~R.~Yakovlev,
G. Karczewski, T. Wojtowicz, D. Schuh,  W. Wegscheider,  and S.~D.~Ganichev,
Appl. Phys. Lett. \textbf{97}, 182107 (2010).

\bibitem{footnote2} Note, that comparing our MPGE data obtained at room temperature
with the $g*$-factor behavior measured at low temperatures, 
we do not consider the weak temperature  dependence 
$g^*$-factor,~\protect \cite{Oestr95,Zawadzki08} which may result in a small shift 
of its zero crossing. 

\bibitem{Oestr95}
M. Oestreich and W. W. R\"uhle, Phys. Rev. Lett. {\bf 74},
2315 (1995).

\bibitem{Zawadzki08}W. Zawadzki, P. Pfeffer, R.~Bratschitsch, Z. Chen,  
S.~T. Cundiff, B.~N. Murdin, and C.~R. Pidgeon, Phys. Rev. B \textbf{78},
245203 (2008)

\bibitem{MPGE2005} V.~V.~Bel'kov, S.~D.~Ganichev,  E.~L.~Ivchenko, S.~A.~Tarasenko,
W.~Weber,  S.~Giglberger,  M.~Olteanu,
H.-P.~Tranitz, S.~N.~Danilov, Petra~Schneider,
W.~Wegscheider, D.~Weiss, and W.~Prettl,
J. Phys. C: Condens. Matter  {\bf 17}, 3405 (2005).

\bibitem{footnote3} Note, that, like in the 
discussed above case of the linear MPGE,  the current in the 4~nm broad QW is
enhanced due to the larger parameter $\chi$ compared to the other samples.

\bibitem{footnote4} Here and latter the calculations are 
limited to rather narrow quantum wells, for which only one 
subband is occupied. In a wider QWs other subbands should be 
taken into account which is out of scope of this paper.

\bibitem{footnote5}  We note, that recently an additional possible root of the spin-galvanic effect
has been addressed in Ref.~\protect \onlinecite{GolubJETPLett2007}. Instead of SGE due to band spin splitting considered in Ref.~\onlinecite{Nature02}, it is based 
on the spin-dependent scattering given by Eq.~\protect \eqref{k_sigma}. 
However, this mechanism also yields a contribution proportional to the Zeeman splitting and
does not qualitatively change the discussion of the MPGE variation upon the QW width change.

\bibitem{GolubJETPLett2007} L.~E. Golub, Pis'ma v ZhETF \textbf{85}, 479 (2007) [JETP Lett. \textbf{85}, 393 (2007)]. 


\end{thebibliography}
\end{document}